\documentclass{eptcs}
\providecommand{\event}{VPT 2015} 

\usepackage{pgf}
\usepackage{pgfnodes}
\usepackage{latexsym}
\usepackage{amsmath}
\usepackage{amssymb}
\usepackage{wasysym}
\usepackage{proof}

\newcommand{\expr}[1]{#1{}}
\newcommand{\cont}[2]{#1{#2}}
\newcommand{\var}[2]{\mathit{#1}#2}
\newcommand{\abs}[3]{\lambda #1{.#2{#3}}}
\newcommand{\app}[3]{#1{~#2{#3}}}
\newcommand{\letexp}[4]{{\bf let} ~ #1{ =  #2{~ {\bf in} ~ #3{#4}}}}

\newcommand{\args}[3]{#1{\ldots#2{#3}}}
\newcommand{\fundef}[2]{#1 & = & \expr{#2}}

\newcommand{\cas}[6]{\begin{array}[t]{@{\hspace*{0mm}}l@{\hspace*{1mm}}l@{\hspace*{1mm}}c@{\hspace*{1mm}}l@{\hspace*{0mm}}} 
\multicolumn{4}{@{\hspace*{0mm}}l@{\hspace*{0mm}}}{{\bf case} ~ #1 ~ {\bf of}} \\
& #2{} & \rightarrow & #3{} \\
~~~ | & #4{} & \rightarrow & #5{#6}\end{array}}
\newcommand{\longcas}[8]{\begin{array}[t]{@{\hspace*{0mm}}l@{\hspace*{1mm}}l@{\hspace*{1mm}}c@{\hspace*{1mm}}l@{\hspace*{0mm}}} 
\multicolumn{4}{@{\hspace*{0mm}}l@{\hspace*{0mm}}}{{\bf case} ~ #1 ~ {\bf of}} \\
& #2{} & \rightarrow & #3{} \\
~~~ | & #4{} & \rightarrow & #5{} \\
~~~ | & #6{} & \rightarrow & #7{#8}\end{array}}
\newcommand{\bigcas}[7]{\begin{array}[t]{@{\hspace*{0mm}}l@{\hspace*{1mm}}l@{\hspace*{1mm}}c@{\hspace*{1mm}}l@{\hspace*{0mm}}} 
\multicolumn{4}{@{\hspace*{0mm}}l@{\hspace*{0mm}}}{{\bf case} ~ e ~{\bf of}} \\
& Request_1 & \rightarrow &  #1{} \\
~~~ | & Request_2 & \rightarrow &  #2{} \\
~~~ | & Take_1 & \rightarrow &  #3{} \\
~~~ | & Take_2 & \rightarrow &  #4{} \\
~~~ | & Release_1 & \rightarrow &  #5{} \\
~~~ | & Release_2 & \rightarrow & #6{#7}\end{array}}
\newcommand{\smallcas}[4]{\begin{array}[t]{@{\hspace*{0mm}}l@{\hspace*{1mm}}l@{\hspace*{1mm}}c@{\hspace*{1mm}}l@{\hspace*{0mm}}}
\multicolumn{4}{@{\hspace*{0mm}}l@{\hspace*{0mm}}}{{\bf case} ~ #1 ~ {\bf of}} \\ ~~~~~ #2 ~ \rightarrow ~  #3{#4}\end{array}}
\newcommand{\casedots}[6]{{\bf case}~#1{~{\bf of}~#2{\rightarrow #3{~| 
\cdots|~#4{\rightarrow #5{#6}}}}}}
\newcommand{\where}[3]{\!\!\!\begin{array}[t]{@{\hspace*{0mm}}l@{\hspace*{1mm}}c@{\hspace*{1mm}}l@{\hspace*{0mm}}}\multicolumn{3}{@{\hspace*{0mm}}l@{\hspace*{0mm}}}{#1{}}\\\multicolumn{3}{@{\hspace*{0mm}}l@{\hspace*{0mm}}}{{\bf where}}\\#2{#3}\end{array}}

\newcommand{\longcons}[3]{Cons~\begin{array}[t]{@{\hspace*{0mm}}l@{\hspace*{0mm}}} #1 \\ #2{#3} \end{array}}
\newcommand{\Where}[6]{#1{} ~ {\bf where} ~ #2 = #3 \ldots #4 = #5{#6}}
\newcommand{\brackets}[2]{(#1{)#2}}

\newcommand{\prove}[5]{{\cal P}[\![#1{]\!]~#2{~#3{~#4{#5}}}}}

\newcommand{\wildcard}[0]{\underline{\hspace{2mm}}}
\newcommand{\ignore}[1]{}

\newtheorem{theorem}{Theorem}[section]

\newtheorem{definition}[theorem]{Definition}  
\newtheorem{example}{Example}  
\newtheorem{property}{Property}  

\title{Verifying Temporal Properties of Reactive Systems by Transformation}

\author{
G.W. Hamilton
\institute{School of Computing and Lero\\
Dublin City University\\
Ireland}
\email{hamilton@computing.dcu.ie}
}

\def\titlerunning{Verifying Temporal Properties by Transformation}

\def\authorrunning{G.W. Hamilton}

\begin{document}

\maketitle

\begin{abstract}
We show how program transformation techniques can be used for the verification of both safety and liveness
properties of reactive systems. In particular, we show how the program transformation technique {\em distillation}
can be used to transform reactive systems specified in a functional language into a simplified form that can subsequently 
be analysed to verify temporal properties of the systems. Example systems which are intended to model mutual exclusion 
are analysed using these techniques with respect to both safety (mutual exclusion) and liveness (non-starvation), with the 
errors they contain being correctly identified.
\end{abstract}

\section{Introduction}

Formal verification of software components is gaining more and more prominence as a viable methodology for
increasing the reliability and reducing the cost of software production. We consider here the problem of verifying 
properties of {\em reactive systems}, i.e., systems which continuously react to external events by changing their 
internal state and producing outputs.  The properties of such systems are usually expressed using a {\em temporal
logic} such as Computational Tree Logic (CTL) or Linear-time Temporal Logic (LTL). These logics are used to
express {\em safety} properties which essentially state that nothing bad will happen, and {\em liveness} properties
which essentially state that something good will eventually happen.

Model checking is a well established technique originally developed for the verification of temporal properties of 
finite state systems \cite{CLARKE86}. However, reactive systems usually have an infinite number of states.
Model checking techniques therefore need to be extended to handle such systems, but the problem of verifying 
such systems is undecidable in general. Most proposed approaches to this problem are semi-automatic and
involve either mathematical {\em (co-)induction} \cite{PNUELI96,SIPMA99} or {\em abstraction} to finite state models 
\cite{ZUCK04,ABDULLA09}. Fold/unfold program transformation techniques have more recently been proposed 
as an automatic approach to this problem. Folding corresponds to the application of a (co-)inductive hypothesis
and generalisation corresponds to abstraction. Many such techniques have been developed for logic programs (e.g. 
\cite{LEUSCHEL99,ROYCHOUDHURI00,FIORAVANTI01,PETTOROSSI09,SEKI11}). However, very few such techniques 
have been developed for functional programs (with the work of Lisitsa and Nemytykh \cite{LISITSA07,LISITSA08} using
supercompilation \cite{TURCHIN86} being a notable exception), and these deal only with safety properties.

In this paper, we show how a fold/unfold program transformation technique can be used to facilitate the verification of both safety 
and liveness properties of reactive systems which have been specified using a functional language. The program transformation
technique which we use is our own {\em distillation} \cite{HAMILTON07A,HAMILTON12} which builds on top of positive 
supercompilation \cite{SORENSEN96}, but is much more powerful. Distillation is used to transform programs defining reactive 
systems into a simplified form which makes them much easier to analyse. We then define a number of verification rules on this 
simplified form to verify temporal properties of the system. In these verification rules, intermediate structures are given an 
undefined value, thus abstracting the system to a finite number of states, but leading to a loss of information.  We argue that, 
since distillation removes more intermediate structures than positive supercompilation, more accurate results are obtained. 
The described techniques are applied to a number of example systems which are intended to model mutually exclusive access 
to a critical resource by two processes, revealing a number of errors.

The remainder of this paper is structured as follows. In Section 2, we introduce the functional language over which our
verification techniques are defined. In Section 3, we show how to specify reactive systems in our language, 
and give a number of example systems which are intended to model mutually exclusive access to a critical resource by two 
processes. In Section 4, we describe how to specify temporal properties for reactive systems defined in our language, and
specify both safety (mutual exclusion) and  liveness (non-starvation) for the example systems. In Section 5, we describe our 
technique for verifying temporal properties of reactive systems and apply this technique to the example systems to verify the 
previously specified temporal properties. Section 6 concludes and considers related work.

\section{Language}
 \label{sec-language-definition}

In this section, we describe the syntax and semantics of the higher-order functional language which will be used 
throughout this paper. 

\subsection{Syntax}

The syntax of our language is given in Figure \ref{grammar}.
\begin{figure}[htb]
\begin{center}
\begin{tabular}{@{\hspace*{0mm}}l@{\hspace*{1mm}}r@{\hspace*{1mm}}l@{\hspace*{1mm}}l@{\hspace*{0mm}}}
$\expr{\var{e}}$ & ::= & $\expr{\var{x}}$ & Variable \\
& $|$ & $\expr{\app{\var{c}}{\args{\var{e_1}}{\var{e_k}}}}$ & Constructor Application \\
& $|$ & $\expr{\abs{\var{x}}{\var{e}}}$ & $\lambda$-Abstraction \\
& $|$ & $\expr{\var{f}}$ & Function Call \\
& $|$ & $\expr{\app{\var{e_0}}{\var{e_1}}}$ & Application \\
& $|$ & $\expr{\casedots{\var{e_0}}{\var{p_1}}{\var{e_1}}{\var{p_k}}{\var{e_k}}}$ & Case Expression \\ 
& $|$ & $\expr{\letexp{\var{x}}{\var{e_0}}{\var{e_1}}}$ & Let Expression \\
& $|$ & $\expr{\Where{\var{e_0}}{\var{f_1}}{\var{e_1}}{\var{f_n}}{\var{e_n}}}$ & Local Function Definitions \\
\\
$\expr{\var{p}}$ & ::= & $\expr{\app{\var{c}}{\args{\var{x_1}}{\var{x_k}}}}$ & Pattern
\end{tabular}
\end{center}
\caption{Language Grammar}
\label{grammar}
\end{figure} \\
A program in the language is an expression which can be a variable, constructor application, $\lambda$-abstraction, 
function call, application, {\bf case}, {\bf let} or {\bf where}. 
Variables introduced by $\lambda$-abstractions, {\bf let} expressions and {\bf case} patterns are {\em bound}; all other variables 
are {\em free}. An expression which contains no free variables is said to be {\em closed}.

Each constructor has a fixed arity; for example $\expr{\var{Nil}}$ has arity 0
and $\expr{\var{Cons}}$ has arity 2.  In an expression $\expr{\app{\var{c}}{\args{\var{e_{1}}}{\var{e_{n}}}}}$,
$n$ must equal the arity of $c$. 
The patterns in {\bf case} expressions may not be nested.  No variable may appear more than once within a pattern. 
We assume that the patterns in a {\bf case} expression are non-overlapping and exhaustive. We also allow a wildcard pattern
$\wildcard$ which always matches if none of the earlier patterns match. Types are defined using algebraic data types, and it is 
assumed that programs are well-typed. Erroneous terms such as $\expr{\casedots{\brackets{\abs{x}{e}}}{\var{p_1}}{\var{e_1}}{\var{p_k}}{\var{e_k}}}$ and $\expr{\app{\brackets{\app{\var{c}}{\args{\var{e_1}}{\var{e_n}}}}}{\var{e}}}$ where $c$ is of arity $n$ cannot therefore occur.

\subsection{Semantics}

The call-by-name operational semantics of our language is standard: we define an evaluation relation $\Downarrow$ between 
closed expressions and {\em values}, where values are expressions in {\em weak head normal form} (i.e. constructor applications or $\lambda$-abstractions). 
We define a one-step reduction relation $\overset{r}{\leadsto}$ inductively as shown in Figure \ref{reduction}, where the reduction $r$ can be $f$ (unfolding of function $f$), $c$ (elimination of constructor $c$) or $\beta$ ($\beta$-substitution). 
\begin{figure}[htb]
\begin{center}
\begin{tabular}{c@{\hspace*{1cm}}c}
$((\lambda x.e_0)~e_1) \overset{\beta}{\leadsto} (e_0\{x \mapsto e_1\})$ & $(\expr{\letexp{\var{x}}{\var{e_0}}{\var{e_1}}}) \overset{\beta}{\leadsto} (e_1\{x \mapsto e_0\})$ \\
\\
$\infer{f \overset{f}{\leadsto} e}{f=e}$ & $\infer{(e_0~e_1) \overset{r}{\leadsto} (e_0'~e_1)}{e_0 \overset{r}{\leadsto} e_0'}$  \\
\\
\multicolumn{2}{c}{$\infer{(\mathbf{case}~(c~e_1 \ldots e_n)~\mathbf{of}~p_1:e_1' | \ldots | p_k:e_k') \overset{c}{\leadsto} (e_i\{x_1 \mapsto e_1,\ldots,x_n \mapsto e_n\})}{p_i=c~x_1 \ldots x_n}$} \\
\\
\multicolumn{2}{c}{$\infer{(\mathbf{case}~e_0~\mathbf{of}~p_1:e_1 | \ldots p_k:e_k) \overset{r}{\leadsto} (\mathbf{case}~e_0'~\mathbf{of}~p_1:e_1 | \ldots p_k:e_k)}{e_0 \overset{r}{\leadsto} e_0'}$}
\end{tabular} 
\end{center}
\caption{One-Step Reduction Relation}
\label{reduction}
\end{figure} \\
We use the notation $e\overset{r}{\leadsto}$ if the expression $e$ reduces, $e\!\Uparrow$ if $e$ diverges, $e\!\Downarrow$ if $e$ converges and 
$e\!\Downarrow\!v$ if $e$ evaluates to the value $v$. These are defined as follows, where $\overset{r}{\leadsto}^*$ denotes the reflexive transitive closure of $\overset{r}{\leadsto}$:
\begin{center}
\begin{tabular}{l@{\hspace{0.5cm}}l}
$e \overset{r}{\leadsto}$, iff $\exists e'.e \overset{r}{\leadsto} e'$ & $e\!\Downarrow$, iff $\exists v.e\!\Downarrow\!v$ \\
$e\!\Downarrow\!v$, iff $e \overset{r}{\leadsto}^* v \wedge \neg(v\overset{r}{\leadsto})$ & $e\!\Uparrow$, iff $\forall e'.e \overset{r}{\leadsto}^* \!e' \Rightarrow e'\overset{r}{\ \leadsto\ }$
\end{tabular}
\end{center}
\section{Specifying Reactive Systems}

In this section, we show how to specify reactive systems in our programming language.
While reactive systems are usually specified using {\em labelled transitions systems}, our
specifications can be trivially derived from these.
Reactive systems have to react to a series of {\em external events} by updating their {\em state}.
In order to facilitate this, we make use of a {\em stream} datatype, which is defined as follows:
$$Stream~a ::= Cons~a~Stream$$
A stream is therefore an infinite list of elements of type $a$. 
Our programs will map an input stream of external events and an initial state to an output stream of {\em observable states},
which give the values of a subset of state variables whose properties can be verified.
In this paper, we wish to analyse a number of systems which are intended to implement mutually exclusive
access to a critical resource for two processes. In all of these systems, the external events belong to the following datatype: 
$$Event ::= Request_1~|~Request_2~|~Take_1~|~Take_2~|~Release_1~|~Release_2$$
Each of the two processes can therefore request access to the critical resource, and take and release
this resource. Observable states in all of our example systems belong to the following datatype:
$$State ::= ObsState~ProcState~ProcState$$
$$ProcState ::= T~|~W~|~U$$
Each process can therefore be thinking ($T$), waiting for the critical resource ($W$) or using the critical resource ($U$). 
In all of the following examples, the variable $es$ represents the external event stream, and $s_1$ and $s_2$ represent 
the states of the two processes respectively.
\begin{example}
\normalfont{In the first example shown in Figure \ref{example1}, each process can request access to the critical resource if 
neither process is using it, take the critical resource if it is waiting for it, and release the critical resource if it is using it.}
\end{example}
\begin{figure}[htb]
\begin{tabular}{l}
$\expr{\where{\app{\app{\app{\var{f}}{\var{es}}}{\var{T}}}{\var{T}}}{\fundef{\var{f}}{\abs{\var{es~s_1~s_2}}{\app{\app{\var{Cons}}{\var{(ObsState~s_1~s_2)}}}{\brackets{\smallcas{\var{es}}{\app{\app{\var{Cons}}{\var{e}}}{\var{es}}}{\bigcas
{\cas{\var{s_1}}{\var{U}}{\app{\app{\app{\var{f}}{\var{es}}}{\var{s_1}}}{\var{s_2}}}{\var{\wildcard}}{\cas{\var{s_2}}{\var{U}}{\app{\app{\app{\var{f}}{\var{es}}}{\var{s_1}}}{\var{s_2}}}{\var{\wildcard}}{\app{\app{\app{\var{f}}{\var{es}}}{\var{W}}}{\var{s_2}}}}}
{\cas{\var{s_2}}{\var{U}}{\app{\app{\app{\var{f}}{\var{es}}}{\var{s_1}}}{\var{s_2}}}{\var{\wildcard}}{\cas{\var{s_1}}{\var{U}}{\app{\app{\app{\var{f}}{\var{es}}}{\var{s_1}}}{\var{s_2}}}{\var{\wildcard}}{\app{\app{\app{\var{f}}{\var{es}}}{\var{s_1}}}{\var{W}}}}}
{\cas{\var{s_1}}{\var{W}}{\app{\app{\app{\var{f}}{\var{es}}}{\var{U}}}{\var{s_2}}}{\var{\wildcard}}{\app{\app{\app{\var{f}}{\var{es}}}{\var{s_1}}}{\var{s_2}}}}
{\cas{\var{s_2}}{\var{W}}{\app{\app{\app{\var{f}}{\var{es}}}{\var{s_1}}}{\var{U}}}{\var{\wildcard}}{\app{\app{\app{\var{f}}{\var{es}}}{\var{s_1}}}{\var{s_2}}}}
{\cas{\var{s_1}}{\var{U}}{\app{\app{\app{\var{f}}{\var{es}}}{\var{T}}}{\var{s_2}}}{\var{\wildcard}}{\app{\app{\app{\var{f}}{\var{es}}}{\var{s_1}}}{\var{s_2}}}}
{\cas{\var{s_2}}{\var{U}}{\app{\app{\app{\var{f}}{\var{es}}}{\var{s_1}}}{\var{T}}}{\var{\wildcard}}{\app{\app{\app{\var{f}}{\var{es}}}{\var{s_1}}}{\var{s_2}}}}}}}}}}}$
\end{tabular}
\caption{Example 1}
\label{example1}
\end{figure}

\begin{example}
\normalfont{In the second example shown in Figure \ref{example2}, each process can request access to the critical resource if it is thinking, take the critical resource if it is waiting for it and the other process is thinking, and release the critical resource if it is using it.}
\end{example}
\begin{figure}[htb]
\begin{tabular}{l}
$\expr{\where{\app{\app{\app{\var{f}}{\var{es}}}{\var{T}}}{\var{T}}}{\fundef{\var{f}}{\abs{\var{es~s_1~s_2}}{\app{\app{\var{Cons}}{\var{(ObsState~s_1~s_2)}}}{\brackets{\smallcas{\var{es}}{\app{\app{\var{Cons}}{\var{e}}}{\var{es}}}{\bigcas
{\cas{\var{s_1}}{\var{T}}{\app{\app{\app{\var{f}}{\var{es}}}{\var{W}}}{\var{s_2}}}{\var{\wildcard}}{\app{\app{\app{\var{f}}{\var{es}}}{\var{s_1}}}{\var{s_2}}}}
{\cas{\var{s_2}}{\var{T}}{\app{\app{\app{\var{f}}{\var{es}}}{\var{s_1}}}{\var{W}}}{\var{\wildcard}}{\app{\app{\app{\var{f}}{\var{es}}}{\var{s_1}}}{\var{s_2}}}}
{\cas{\var{s_1}}{\var{W}}{\cas{\var{s_2}}{\var{T}}{\app{\app{\app{\var{f}}{\var{es}}}{\var{U}}}{\var{s_2}}}{\var{\wildcard}}{\app{\app{\app{\var{f}}{\var{es}}}{\var{s_1}}}{\var{s_2}}}}{\var{\wildcard}}{\app{\app{\app{\var{f}}{\var{es}}}{\var{s_1}}}{\var{s_2}}}}
{\cas{\var{s_2}}{\var{W}}{\cas{\var{s_1}}{\var{T}}{\app{\app{\app{\var{f}}{\var{es}}}{\var{s_1}}}{\var{U}}}{\var{\wildcard}}{\app{\app{\app{\var{f}}{\var{es}}}{\var{s_1}}}{\var{s_2}}}}{\var{\wildcard}}{\app{\app{\app{\var{f}}{\var{es}}}{\var{s_1}}}{\var{s_2}}}}
{\cas{\var{s_1}}{\var{U}}{\app{\app{\app{\var{f}}{\var{es}}}{\var{T}}}{\var{s_2}}}{\var{\wildcard}}{\app{\app{\app{\var{f}}{\var{es}}}{\var{s_1}}}{\var{s_2}}}}
{\cas{\var{s_2}}{\var{U}}{\app{\app{\app{\var{f}}{\var{es}}}{\var{s_1}}}{\var{T}}}{\var{\wildcard}}{\app{\app{\app{\var{f}}{\var{es}}}{\var{s_1}}}{\var{s_2}}}}}}}}}}}$
\end{tabular} 
\caption{Example 2}
\label{example2}
\end{figure}

\begin{example}
\normalfont{In the final example in Figure \ref{example3}, we implement Lamport's bakery algorithm \cite{LAMPORT74} 
for two processes. In this example, to request access to the critical resource, each process must take a `ticket' with a number, 
and the process with the lowest valued ticket is given precedence. A ticket value of zero indicates that a process has not 
requested to use the critical resource, so when a process releases the critical resource its ticket value is reset to zero. 
We therefore add two state variables $t_1$ and $t_2$ which give the current ticket number for each process, but are
not part of the observable state. These are natural numbers belonging to the following datatype:
$$Nat ::= Zero~|~Succ~Nat$$
Note that, since there is no limit to the number of a ticket (ticket numbers will keep increasing if each process alternately requests access to the critical resource before the other process has released it), this is an example of an infinite state system which can cause 
problems for some model checkers.}
\end{example}
\begin{figure}[htb]
\begin{tabular}{l}
$\expr{\where{\app{\app{\app{\app{\app{\var{f}}{\var{es}}}{\var{T}}}{\var{T}}}{\var{Zero}}}{\var{Zero}}}{\fundef{\var{f}}{\abs{\var{es~s_1~s_2~t_1~t_2}}{\longcons{\var{(ObsState~s_1~s_2)}}{\brackets{\smallcas{\var{es}}{\app{\app{\var{Cons}}{\var{e}}}{\var{es}}}{\bigcas
{\cas{\var{s_1}}{\var{T}}{\app{\app{\app{\app{\app{\var{f}}{\var{es}}}{\var{W}}}{\var{s_2}}}{\brackets{\app{\var{Succ}}{\var{t_2}}}}}{\var{t_2}}}{\var{\wildcard}}{\app{\app{\app{\app{\app{\var{f}}{\var{es}}}{\var{s_1}}}{\var{s_2}}}{\var{t_1}}}{\var{t_2}}}}
{\cas{\var{s_2}}{\var{T}}{\app{\app{\app{\app{\app{\var{f}}{\var{es}}}{\var{s_1}}}{\var{W}}}{\var{t_1}}}{\brackets{\app{\var{Succ}}{\var{t_1}}}}}{\var{\wildcard}}{\app{\app{\app{\app{\app{\var{f}}{\var{es}}}{\var{s_1}}}{\var{s_2}}}{\var{t_1}}}{\var{t_2}}}}
{\cas{\var{s_1}}{\var{W}}{\cas{\var{s_2}}{\var{T}}{\app{\app{\app{\app{\app{\var{f}}{\var{es}}}{\var{U}}}{\var{s_2}}}{\var{t_1}}}{\var{t_2}}}{\var{\wildcard}}{\cas{\brackets{\var{t_1 < t_2}}}{\var{True}}{\app{\app{\app{\app{\app{\var{f}}{\var{es}}}{\var{U}}}{\var{s_2}}}{\var{t_1}}}{\var{t_2}}}{\var{False}}{\app{\app{\app{\app{\app{\var{f}}{\var{es}}}{\var{s_1}}}{\var{s_2}}}{\var{t_1}}}{\var{t_2}}}}}{\var{\wildcard}}{\app{\app{\app{\app{\app{\var{f}}{\var{es}}}{\var{s_1}}}{\var{s_2}}}{\var{t_1}}}{\var{t_2}}}}
{\cas{\var{s_2}}{\var{W}}{\cas{\var{s_1}}{\var{T}}{\app{\app{\app{\app{\app{\var{f}}{\var{es}}}{\var{s_1}}}{\var{U}}}{\var{t_1}}}{\var{t_2}}}{\var{\wildcard}}{\cas{\brackets{\var{t_2 < t_1}}}{\var{True}}{\app{\app{\app{\app{\app{\var{f}}{\var{es}}}{\var{s_1}}}{\var{U}}}{\var{t_1}}}{\var{t_2}}}{\var{False}}{\app{\app{\app{\app{\app{\var{f}}{\var{es}}}{\var{s_1}}}{\var{s_2}}}{\var{t_1}}}{\var{t_2}}}}}{\var{\wildcard}}{\app{\app{\app{\app{\app{\var{f}}{\var{es}}}{\var{s_1}}}{\var{s_2}}}{\var{t_1}}}{\var{t_2}}}}
{\cas{\var{s_1}}{\var{U}}{\app{\app{\app{\app{\app{\var{f}}{\var{es}}}{\var{T}}}{\var{s_2}}}{\var{Zero}}}{\var{t_2}}}{\var{\wildcard}}{\app{\app{\app{\app{\app{\var{f}}{\var{es}}}{\var{s_1}}}{\var{s_2}}}{\var{t_1}}}{\var{t_2}}}}
{\cas{\var{s_2}}{\var{U}}{\app{\app{\app{\app{\app{\var{f}}{\var{es}}}{\var{s_1}}}{\var{T}}}{\var{t_1}}}{\var{Zero}}}{\var{\wildcard}}{\app{\app{\app{\app{\app{\var{f}}{\var{es}}}{\var{s_1}}}{\var{s_2}}}{\var{t_1}}}{\var{t_2}}}}}}}}}}}$
\end{tabular} 
\caption{Example 3}
\label{example3}
\end{figure}
\section{Specification of Temporal Properties}

In this section, we describe how temporal properties of reactive systems defined in our functional language are specified.
We use Linear-time Temporal Logic (LTL), in which the set of well-founded formulae (WFF) 
are defined inductively as follows. All atomic propositions $p$ are in WFF; if $\varphi$ and $\psi$ are in WFF, then so are:
\begin{itemize}
\item $\neg \varphi$
\item $\varphi \vee \psi$
\item $\varphi \wedge \psi$
\item $\varphi \Rightarrow \psi$
\item $\Box \varphi$
\item $\Diamond \varphi$
\item $\ocircle \varphi$
\end{itemize}
The temporal operator $\Box \varphi$ means that $\varphi$ is {\em always} true; this is used to express {\em safety}
properties. The temporal operator $\Diamond \varphi$ means that $\varphi$ will {\em eventually} be true; this is used to 
express {\em liveness} properties. The temporal operator $\ocircle \varphi$ means that $\varphi$ is true in the {\em next} state.
These modalities can be combined to obtain new modalities; for example, $\Box \Diamond \varphi$ means that $\varphi$ 
is true infinitely often, and $\Diamond \Box \varphi$ means that $\varphi$ is eventually true forever. 
Fairness constraints can also be specified for some external events (those belonging to the set $F$) which require that they occur infinitely often. For the examples given in this paper, it is assumed that all external events belong to $F$.

Propositional models for linear-time temporal formulas consist of an infinite sequence of states 
$\pi = \langle s_0, s_1, \ldots \rangle$ such that each state $s_i$ supplies an assignment to the atomic propositions. 
The satisfaction relation is extended to formulas in LTL for a model $\pi$ and position $i$ as follows. 
\begin{center}
\begin{tabular}{lcl}
$\pi,i \vDash p$ & iff & $p \in s_i$ \\
$\pi,i \vDash \neg \varphi$ & iff & $\pi,i \nvDash \varphi$ \\
$\pi,i \vDash \varphi \vee \psi$ & iff & $\pi,i \vDash \varphi$ or $\pi,i \vDash \psi$ \\
$\pi,i \vDash \varphi \wedge \psi$ & iff & $\pi,i \vDash \varphi$ and $\pi,i \vDash \psi$ \\
$\pi,i \vDash \varphi \Rightarrow \psi$ & iff & $\pi,i \nvDash \varphi$ or $\pi,i \vDash \psi$ \\
$\pi,i \vDash \Box \varphi$ & iff & $\forall j \geq i.\pi,j \vDash \varphi$ \\
$\pi,i \vDash \Diamond \varphi$ & iff & $\exists j \geq i.\pi,j \vDash \varphi$ \\
$\pi,i \vDash \ocircle \varphi$ & iff & $\pi,i+1 \vDash \varphi$
\end{tabular}
\end{center}
A formula $\varphi$ holds in model $\pi$ if it holds at position 0 i.e. $\pi,0 \vDash \varphi$.

The atomic propositions of these temporal formulae can be trivially translated into our functional language.
For our verification rules, we define the following datatype for truth values:
$$TruthVal ::= \mathit{True}~|~\mathit{False}~|~\mathit{Undefined}$$
We use a Kleene three-valued logic because our verification rules must always return an answer, but some of the properties to be verified may be undecidable.
For our example programs which attempt to implement mutual exclusion, the following two properties are defined.
Within these temporal properties, we use the variable $s$ to denote the current observable state whose properties are being specified.
\begin{property}[Mutual Exclusion]
\normalfont{This is a safety property which specifies that both processes cannot be using the critical resource at the same time.
This can be specified as follows: \\
\hspace*{4cm} $\Box \expr{\brackets{\smallcas{\var{s}}{\var{ObsState~s_1~s_2}}{\cas{\var{s_1}}{\var{U}}{\cas{\var{s_2}}{\var{U}}{\var{False}}{\var{\wildcard}}{\var{True}}}{\var{\wildcard}}{\var{True}}}}}$
}
\end{property}
\begin{property}[Non-Starvation]
\normalfont{This is a liveness property which specifies that each process must eventually get to use the critical resource 
if they are waiting for it. This can be specified for process 1 as follows: \\
\hspace*{1.5cm} $\cont{\Box (\brackets{\smallcas{\var{s}}{\var{ObsState~s_1~s_2}}{\cas{\var{s_1}}{\var{W}}{\var{True}}{\var{\wildcard}}{\var{False}}}}}
{\Rightarrow \Diamond \expr{\brackets{\smallcas{\var{s}}{\var{ObsState~s_1~s_2}}{\cas{\var{s_1}}{\var{U}}{\var{True}}{\var{\wildcard}}{\var{False}}}})}}$ \\
The specification of this property for process 2 is similar. 
}
\end{property}

\section{Verification of Temporal Properties}

In this section, we show how temporal properties of reactive systems defined in our functional language can be verified.
To facilitate this, we first of all transform the reactive systems definitions into a simplified form using distillation 
\cite{HAMILTON07A,HAMILTON12}, a powerful program transformation technique which builds on top
of the supercompilation transformation \cite{TURCHIN86,SORENSEN96}. Due to the nature of the reactive systems definitions,
in which the input is an external event stream, and the output is a stream of observable states, the programs resulting from
this transformation will take the form $e^{\emptyset}$, where $e^{\rho}$ is defined as follows.
\begin{center}
\begin{tabular}{lrl}
$\expr{\var{e^{\rho}}}$ & ::= & $\expr{\app{\app{\var{Cons}}{\var{e_0^{\rho}}}}{\var{e_1^{\rho}}}}$ \\
& $|$ & $\expr{\app{\var{f}}{\args{\var{x_1}}{\var{x_n}}}}$ \\
& $|$ & $\expr{\casedots{\var{x}}{\var{p_1}}{\var{e_1^{\rho}}}{\var{p_k}}{\var{e_n^{\rho}}}}$, where $x \notin \rho$ \\ 
& $|$ & $\expr{\app{\var{x}}{\args{\var{e_1^{\rho}}}{\var{e_n^{\rho}}}}}$, where $x \in \rho$ \\
& $|$ & $\expr{\letexp{\var{x}}{\abs{\args{\var{x_1}}{\var{x_n}}}{\var{e_0^{\rho}}}}{\var{e_1^{(\rho \cup \{x\})}}}}$ \\
& $|$ & $\expr{\Where{\var{e_0^{\rho}}}{\var{f_1}}{\abs{\args{\var{x_{1_1}}}{\var{x_{1_k}}}}{\var{e_1^{\rho}}}}{\var{f_n}}{\abs{\args{\var{x_{n_1}}}{\var{x_{n_k}}}}{\var{e_n^{\rho}}}}}$
\end{tabular} 
\end{center}
The {\bf let} variables are added to the set $\rho$,
and will not be used in the selectors of {\bf case} expressions. These {\bf let} variables are given an undefined value during 
verification, thus abstracting the system to a finite number of states. 

We define our verification rules on this restricted form of program as shown in Figure \ref{proofrules}.
\begin{figure}[htbp]
\begin{center}
\begin{tabular}[t]{@{\hspace*{0mm}}l@{\hspace*{1mm}}l@{\hspace*{1mm}}c@{\hspace*{1mm}}l@{\hspace*{0mm}}}
(1) & $\expr{\prove{\var{e}}{(\varphi \wedge \psi)}{\phi}{\rho}}$ &  = & $\expr{\longcas{\brackets{\prove{\var{e}}{\varphi}{\phi}{\rho}}}{\var{True}}{\prove{\var{e}}{\psi}{\phi}{\rho}}{\var{False}}{\var{False}}{\var{\mathit{Undefined}}}{\cas{\brackets{\prove{\var{e}}{\psi}{\phi}{\rho}}}{\var{False}}{\var{False}}{\var{\wildcard}}{\var{\mathit{Undefined}}}}}$ \\
(2) & $\expr{\prove{\var{e}}{(\varphi \vee \psi)}{\phi}{\rho}}$ &  = & $\expr{\longcas{\brackets{\prove{\var{e}}{\varphi}{\phi}{\rho}}}{\var{True}}{\var{True}}{\var{False}}{\prove{\var{e}}{\psi}{\phi}{\rho}}{\var{\mathit{Undefined}}}{\cas{\brackets{\prove{\var{e}}{\psi}{\phi}{\rho}}}{\var{True}}{\var{True}}{\var{\wildcard}}{\var{\mathit{Undefined}}}}}$ \\
(3) & $\expr{\prove{\var{e}}{(\varphi \Rightarrow \psi)}{\phi}{\rho}}$ &  = & $\expr{\longcas{\brackets{\prove{\var{e}}{\varphi}{\phi}{\rho}}}{\var{True}}{\prove{\var{e}}{\psi}{\phi}{\rho}}{\var{False}}{\var{True}}{\var{\mathit{Undefined}}}{\cas{\brackets{\prove{\var{e}}{\psi}{\phi}{\rho}}}{\var{True}}{\var{True}}{\var{\wildcard}}{\var{\mathit{Undefined}}}}}$ \\
(4) & $\expr{\prove{\var{e}}{(\neg \varphi)}{\phi}{\rho}}$ &  = & $\expr{\longcas{\brackets{\prove{\var{e}}{\varphi}{\phi}{\rho}}}{\var{True}}{\var{False}}{\var{False}}{\var{True}}{\var{\mathit{Undefined}}}{\var{\mathit{Undefined}}}}$ \\
(5a) & $\expr{\prove{\app{\app{\var{Cons}}{\var{e_0}}}{\var{e_1}}}{(\Box \varphi)}{\phi}{\rho}}$ &  = & $\expr{\prove{\app{\app{\var{Cons}}{\var{e_0}}}{\var{e_1}}}{\varphi}{\phi}{\emptyset}} \wedge 
\expr{\prove{\var{e_1}}{(\Box \varphi)}{\phi}{\rho}}$ \\
(5b) & $\expr{\prove{\app{\app{\var{Cons}}{\var{e_0}}}{\var{e_1}}}{(\Diamond \varphi)}{\phi}{\rho}}$ &  = & $\expr{\prove{\app{\app{\var{Cons}}{\var{e_0}}}{\var{e_1}}}{\varphi}{\phi}{\emptyset}} \vee 
\expr{\prove{\var{e_1}}{(\Diamond \varphi)}{\phi}{\rho}}$ \\
(5c) & $\expr{\prove{\app{\app{\var{Cons}}{\var{e_0}}}{\var{e_1}}}{(\ocircle \varphi)}{\phi}{\rho}}$ &  = & $\expr{\prove{\var{e_1}}{\varphi}{\phi}{\rho}}$ \\
(5d) & $\expr{\prove{\app{\app{\var{Cons}}{\var{e_0}}}{\var{e_1}}}{\varphi}{\phi}{\rho}}$ &  = & $v$, where $\varphi[e_0/s] \Downarrow v$ \\
(6a) & $\expr{\prove{\app{\var{f}}{\args{x_1}{x_n}}}{(\Box \varphi)}{\phi}{\rho}}$ & = & $\left\{\begin{tabular}[c]{@{\hspace*{0mm}}l@{\hspace*{1mm}}l@{\hspace*{0mm}}}
$\expr{\var{True}}$, & if $f \in \rho$ \\
$\expr{\prove{\var{e}[x_1/x_1',\ldots,x_n/x_n']}{(\Box \varphi)}{\phi}{(\rho \cup \{f\})}}$, & otherwise 
\end{tabular}\right.$ \\
& & & where $\phi(f) = \expr{\abs{\args{\var{x_1'}}{\var{x_n'}}}{\var{e}}}$ \\
(6b) & $\expr{\prove{\app{\var{f}}{\args{x_1}{x_n}}}{(\Diamond \varphi)}{\phi}{\rho}}$ & = & $\left\{\begin{tabular}[c]{@{\hspace*{0mm}}l@{\hspace*{1mm}}l@{\hspace*{0mm}}}
$\expr{\var{False}}$, & if $f \in \rho$ \\
$\expr{\prove{\var{e}[x_1/x_1',\ldots,x_n/x_n']}{(\Diamond \varphi)}{\phi}{(\rho \cup \{f\})}}$, & otherwise 
\end{tabular}\right.$ \\
& & & where $\phi(f) = \expr{\abs{\args{\var{x_1'}}{\var{x_n'}}}{\var{e}}}$ \\
(6c) & $\expr{\prove{\app{\var{f}}{\args{x_1}{x_n}}}{\varphi}{\phi}{\rho}}$ & = & 
$\left\{\begin{tabular}[c]{@{\hspace*{0mm}}l@{\hspace*{1mm}}l@{\hspace*{0mm}}}
$\expr{\var{Undefined}}$, & if $f \in \rho$ \\
$\expr{\prove{\var{e}[x_1/x_1',\ldots,x_n/x_n']}{\varphi}{\phi}{(\rho \cup \{f\})}}$, & otherwise 
\end{tabular}\right.$ \\
& & & where $\phi(f) = \expr{\abs{\args{\var{x_1'}}{\var{x_n'}}}{\var{e}}}$ \\
(7a) & \multicolumn{3}{@{\hspace*{0mm}}l@{\hspace*{0mm}}}{$\expr{\prove{\casedots{\var{x}}{\var{p_1}}{\var{e_1}}{\var{p_n}}{\var{e_n}}}{(\Diamond \varphi)}{\phi}{\rho}}$} \\
& & = & $(\bigvee\limits_{p_i \in F} \expr{\prove{\var{e_i}}{(\Diamond \varphi)}{\phi}{\rho}}) \vee (\bigwedge \limits_{i = 1}^{n}\expr{\prove{\var{e_i}}{(\Diamond \varphi)}{\phi}{\rho}})$ \\
(7b) & \multicolumn{3}{@{\hspace*{0mm}}l@{\hspace*{0mm}}}{$\expr{\prove{\casedots{\var{x}}{\var{p_1}}{\var{e_1}}{\var{p_n}}{\var{e_n}}}{\varphi}{\phi}{\rho}}$} \\
& & = & $\bigwedge\limits_{i=1}^{n} \expr{\prove{\var{e_i}}{\varphi}{\phi}{\rho}}$ \\
(8) & $\expr{\prove{\app{\var{x}}{\args{\var{e_1}}{\var{e_n}}}}{\varphi}{\phi}{\rho}}$ &  = & $\mathit{Undefined}$ \\
(9) & $\expr{\prove{\letexp{\var{x}}{\var{e_0}}{\var{e_1}}}{\varphi}{\phi}{\rho}}$ & = & $\expr{\prove{\var{e_1}}{\varphi}{\phi}{\rho}}$ \\
(10) & \multicolumn{3}{@{\hspace*{0mm}}l@{\hspace*{0mm}}}{$\expr{\prove{\Where{\var{e_0}}{\var{f_1}}{\var{e_1}}{\var{f_n}}{\var{e_n}}}{\varphi}{\phi}{\rho}}$} \\
& & = & $\expr{\prove{\var{e_0}}{\varphi}{(\phi \cup \{f_1 \mapsto e_1,\ldots,f_n \mapsto e_n\})}{\rho}}$
\end{tabular}
\end{center}
\caption{Verification Rules}
\label{proofrules}
\end{figure} 
The parameter $\varphi$ denotes the property to be verified and $\phi$ denotes the function variable environment.
$\rho$ denotes the set of function calls previously encountered; this is used for the detection of loops to ensure termination.
$\rho$ is also used in the verification of the $\Box$ operator (which evaluates to $True$ on encountering a loop), and the 
verification of the $\Diamond$ operator (which evaluates to $False$ on encountering a loop); $\rho$ is reset to empty when 
the verification moves inside these temporal operators. For all other temporal formulae, the value $\mathit{Undefined}$ is 
returned on encountering a loop.

The verification rules can be explained as follows. The logical connectives $\wedge$, $\vee$, $\Rightarrow$ and $\neg$ are 
defined in the usual way for a Kleene three-valued logic in our language in rules (1-4). Rules (5a-d) deal with a constructed stream of 
states. In rule (5a), if we are trying to verify that a property is always true, then we verify that it is true for the first state (with 
$\rho$ reset to empty) and is always true in all remaining states. In rule (5b), if we are trying to verify that a property is 
eventually true, then we verify that it is either true for the first state (with $\rho$ reset to empty) or is eventually true in all 
remaining states. In rule (5c), if we are trying to verify that a property is true in the next state then we verify that the property 
is true for the next state. In rule (5d), if we are trying to verify that a property is true in the current state then we verify that the 
property is true for the current state by evaluating the property using the value of the current state for the state variable $s$. 
Rules (6a-c) deal with function calls. In rule (6a), if we are trying to verify that a property is always true, then if the function 
call has been encountered before while trying to verify the same property we can return the value {\em True}; this corresponds 
to the standard greatest fixed point calculation normally used for the $\Box$ operator in which the property is initially assumed 
to be {\em True} for all states. Otherwise, the function is unfolded and added to the set of previously encountered function calls 
for this property. In rule (6b), if we are trying to verify that a property is eventually true, then if the function call has been 
encountered before while trying to verify the same property we can return the value {\em False}; this corresponds to the 
standard least fixed point calculation normally used for the $\Diamond$ property in which the property is initially assumed to be 
{\em False} for all states. Otherwise, the function is unfolded 
and added to the set of previously encountered function calls for this property. In rule (6c), if we are trying to verify that any 
other property is true, then if the function call has been encountered before we can return the value $\mathit{Undefined}$ since 
a loop has been detected. Otherwise, the function is unfolded and added to the set of previously encountered function calls. 
Rules (7a-b) deal with {\bf case} expressions. In rule (7a), if we are trying to verify that a property is eventually true,
then we verify that it is either eventually true for at least one of the branches for which there is a fairness assumption
(since these branches must be selected eventually), or that it is eventually true for all branches. 
In Rule (7b), if we are trying to verify that any other property is true, then we verify that it is true for all branches. 
In rule (8), if we encounter a free variable, then we return the value $\mathit{Undefined}$ since we cannot determine the value 
of the variable; this must be a {\bf let} variable which has been abstracted, so no information can be determined for it. 
In rule (9), in order to verify that a property is true for a {\bf let} expression, we verify that it is true for the {\bf let} body; 
this is where we perform abstraction of the extracted sub-expression. In rule (10), for a {\bf where} expression, the function 
definitions are added to the environment $\phi$.

\begin{theorem}[Soundness]
\normalfont{$\expr{\prove{e}{\varphi}{\emptyset}{\emptyset}} = True \Rightarrow \pi,0 \vDash \varphi \wedge
\expr{\prove{e}{\varphi}{\emptyset}{\emptyset}} = False \Rightarrow \pi,0 \nvDash \varphi$ \\ where $\pi$ is a model for $e$.}
\end{theorem}
{\bf Proof} \\
The proof of this is by recursion induction on the verification rules ${\cal P}$.
\begin{theorem}[Termination]
\normalfont{$\forall e \in$ Prog, $\varphi \in$ WFF, $\expr{\prove{e}{\varphi}{\emptyset}{\emptyset}}$ always terminates.}
\end{theorem}
{\bf Proof} \\
Proof of termination is quite straightforward since there will be a finite number of functions and uses of the temporal
operators $\Box$ and $\Diamond$, and verification of each of these temporal operators will terminate when a function is
re-encountered. \\
\\
Using these rules, we try to verify the two properties (mutual exclusion and non-starvation) for the example programs 
for mutual exclusion given in Section 3. Firstly, distillation is applied to each of the programs. 
\setcounter{example}{0}
\begin{example}
\normalfont{The result of distilling Example 1 is shown in Figure \ref{example1distilled}, and the LTS representation of this program is shown in Figure \ref{example1lts} (for ease of presentation of this and subsequent LTSs, transitions back into the same state have been omitted).
\begin{figure}[htbp]
\hspace*{2cm}
\begin{tabular}{l}
$\expr{\where{\app{\var{f_1}}{\var{es}}}{
\fundef{\var{f_1}}{\abs{\var{es}}{\app{\app{\var{Cons}}{\var{(ObsState~T~T)}}}{\brackets{\smallcas{\var{es}}{\app{\app{\var{Cons}}{\var{e}}}{\var{es}}}{\longcas{\var{e}}{\var{Request_1}}{\app{\var{f_2}}{\var{es}}}{\var{Request_2}}{\app{\var{f_3}}{\var{es}}}{\var{\wildcard}}{\app{\var{f_1}}{\var{es}}}}}}}} \\
\fundef{\var{f_2}}{\abs{\var{es}}{\app{\app{\var{Cons}}{\var{(ObsState~W~T)}}}{\brackets{\smallcas{\var{es}}{\app{\app{\var{Cons}}{\var{e}}}{\var{es}}}{\longcas{\var{e}}{\var{Take_1}}{\app{\var{f_4}}{\var{es}}}{\var{Request_2}}{\app{\var{f_5}}{\var{es}}}{\var{\wildcard}}{\app{\var{f_2}}{\var{es}}}}}}}} \\
\fundef{\var{f_3}}{\abs{\var{es}}{\app{\app{\var{Cons}}{\var{(ObsState~T~W)}}}{\brackets{\smallcas{\var{es}}{\app{\app{\var{Cons}}{\var{e}}}{\var{es}}}{\longcas{\var{e}}{\var{Request_1}}{\app{\var{f_5}}{\var{es}}}{\var{Take_2}}{\app{\var{f_6}}{\var{es}}}{\var{\wildcard}}{\app{\var{f_3}}{\var{es}}}}}}}} \\
\fundef{\var{f_4}}{\abs{\var{es}}{\app{\app{\var{Cons}}{\var{(ObsState~U~T)}}}{\brackets{\smallcas{\var{es}}{\app{\app{\var{Cons}}{\var{e}}}{\var{es}}}{\cas{\var{e}}{\var{Release_1}}{\app{\var{f_1}}{\var{es}}}{\var{\wildcard}}{\app{\var{f_4}}{\var{es}}}}}}}} \\
\fundef{\var{f_5}}{\abs{\var{es}}{\app{\app{\var{Cons}}{\var{(ObsState~W~W)}}}{\brackets{\smallcas{\var{es}}{\app{\app{\var{Cons}}{\var{e}}}{\var{es}}}{\longcas{\var{e}}{\var{Take_1}}{\app{\var{f_7}}{\var{es}}}{\var{Take_2}}{\app{\var{f_8}}{\var{es}}}{\var{\wildcard}}{\app{\var{f_5}}{\var{es}}}}}}}} \\
\fundef{\var{f_6}}{\abs{\var{es}}{\app{\app{\var{Cons}}{\var{(ObsState~T~U)}}}{\brackets{\smallcas{\var{es}}{\app{\app{\var{Cons}}{\var{e}}}{\var{es}}}{\cas{\var{e}}{\var{Release_2}}{\app{\var{f_1}}{\var{es}}}{\var{\wildcard}}{\app{\var{f_6}}{\var{es}}}}}}}} \\
\fundef{\var{f_7}}{\abs{\var{es}}{\app{\app{\var{Cons}}{\var{(ObsState~U~W)}}}{\brackets{\smallcas{\var{es}}{\app{\app{\var{Cons}}{\var{e}}}{\var{es}}}{\longcas{\var{e}}{\var{Release_1}}{\app{\var{f_3}}{\var{es}}}{\var{Take_2}}{\app{\var{f_9}}{\var{es}}}{\var{\wildcard}}{\app{\var{f_7}}{\var{es}}}}}}}} \\
\fundef{\var{f_8}}{\abs{\var{es}}{\app{\app{\var{Cons}}{\var{(ObsState~W~U)}}}{\brackets{\smallcas{\var{es}}{\app{\app{\var{Cons}}{\var{e}}}{\var{es}}}{\longcas{\var{e}}{\var{Release_2}}{\app{\var{f_2}}{\var{es}}}{\var{Take_1}}{\app{\var{f_9}}{\var{es}}}{\var{\wildcard}}{\app{\var{f_8}}{\var{es}}}}}}}} \\
\fundef{\var{f_9}}{\abs{\var{es}}{\app{\app{\var{Cons}}{\var{(ObsState~U~U)}}}{\brackets{\smallcas{\var{es}}{\app{\app{\var{Cons}}{\var{e}}}{\var{es}}}{\longcas{\var{e}}{\var{Release_1}}{\app{\var{f_6}}{\var{es}}}{\var{Release_2}}{\app{\var{f_4}}{\var{es}}}{\var{\wildcard}}{\app{\var{f_9}}{\var{es}}}}}}}}}}$
\end{tabular}
\caption{Result of Distilling Example 1}
\label{example1distilled}
\end{figure}
\begin{figure}[htb]
\begin{pgfpicture}{0cm}{1cm}{20cm}{10cm}
\pgfnodebox{node1}[stroke]{\pgfxy(8.0,9.5)}{\parbox{1.1cm}{$f_1 \\ s_1 = T \\ s_2 = T$}}{5pt}{5pt}
\pgfnodebox{node2}[stroke]{\pgfxy(5.0,7.5)}{\parbox{1.1cm}{$f_2 \\ s_1 = W \\ s_2 = T$}}{5pt}{5pt}
\pgfsetendarrow{\pgfarrowto}
\pgfnodeconnline{node1}{node2}
\pgfnodelabel{node1}{node2}[0.5][0pt]{\pgfbox[center,center]{$Request_1$}}
\pgfnodebox{node3}[stroke]{\pgfxy(11.0,7.5)}{\parbox{1.1cm}{$f_3 \\ s_1 = T \\ s_2 = W$}}{5pt}{5pt}
\pgfsetendarrow{\pgfarrowto}
\pgfnodeconnline{node1}{node3}
\pgfnodelabel{node1}{node3}[0.5][0pt]{\pgfbox[center,center]{$Request_2$}}
\pgfnodebox{node4}[stroke]{\pgfxy(2.0,5.5)}{\parbox{1.1cm}{$f_4 \\ s_1 = U \\ s_2 = T$}}{5pt}{5pt}
\pgfsetendarrow{\pgfarrowto}
\pgfnodeconnline{node2}{node4}
\pgfnodelabel{node2}{node4}[0.5][0pt]{\pgfbox[center,center]{$Take_1$}}
\pgfsetendarrow{\pgfarrowto}
\pgfnodeconncurve{node4}{node1}{90}{180}{2cm}{2cm}
\pgfnodelabel{node4}{node1}[0.9][1cm]{\pgfbox[center,center]{$Release_1$}}
\pgfnodebox{node5}[stroke]{\pgfxy(8.0,5.5)}{\parbox{1.1cm}{$f_5 \\ s_1 = W \\ s_2 = W$}}{5pt}{5pt}
\pgfsetendarrow{\pgfarrowto}
\pgfnodeconnline{node2}{node5}
\pgfnodelabel{node2}{node5}[0.5][0pt]{\pgfbox[center,center]{$Request_2$}}
\pgfsetendarrow{\pgfarrowto}
\pgfnodeconnline{node3}{node5}
\pgfnodelabel{node3}{node5}[0.5][0pt]{\pgfbox[center,center]{$Request_1$}}
\pgfnodebox{node6}[stroke]{\pgfxy(14.0,5.5)}{\parbox{1.1cm}{$f_6 \\ s_1 = T \\ s_2 = U$}}{5pt}{5pt}
\pgfsetendarrow{\pgfarrowto}
\pgfnodeconnline{node3}{node6}
\pgfnodelabel{node3}{node6}[0.5][0pt]{\pgfbox[center,center]{$Take_2$}}
\pgfsetendarrow{\pgfarrowto}
\pgfnodeconncurve{node6}{node1}{90}{0}{2cm}{2cm}
\pgfnodelabel{node6}{node1}[0.9][-1cm]{\pgfbox[center,center]{$Release_2$}}
\pgfnodebox{node7}[stroke]{\pgfxy(5.0,3.5)}{\parbox{1.1cm}{$f_7 \\ s_1 = U \\ s_2 = W$}}{5pt}{5pt}
\pgfsetendarrow{\pgfarrowto}
\pgfnodeconnline{node5}{node7}
\pgfnodelabel{node5}{node7}[0.5][0pt]{\pgfbox[center,center]{$Take_1$}}
\pgfsetendarrow{\pgfarrowto}
\pgfnodeconncurve{node7}{node1}{180}{180}{6cm}{7cm}
\pgfsetendarrow{\pgfarrowto}
\pgfnodeconncurve{node7}{node3}{0}{270}{2cm}{2cm}
\pgfnodelabel{node7}{node2}[0.8][0cm]{\pgfbox[center,center]{$Release_2$}}
\pgfnodebox{node8}[stroke]{\pgfxy(11.0,3.5)}{\parbox{1.1cm}{$f_8 \\ s_1 = W \\ s_2 = U$}}{5pt}{5pt}
\pgfsetendarrow{\pgfarrowto}
\pgfnodeconnline{node5}{node8}
\pgfnodelabel{node5}{node8}[0.5][0pt]{\pgfbox[center,center]{$Take_2$}}
\pgfsetendarrow{\pgfarrowto}
\pgfnodeconncurve{node8}{node1}{0}{0}{6cm}{7cm}
\pgfnodelabel{node8}{node3}[0.8][0cm]{\pgfbox[center,center]{$Release_1$}}
\pgfsetendarrow{\pgfarrowto}
\pgfnodeconncurve{node8}{node2}{180}{270}{2cm}{2cm}
\pgfnodebox{node9}[stroke]{\pgfxy(8.0,1.5)}{\parbox{1.1cm}{$f_9 \\ s_1 = U \\ s_2 = U$}}{5pt}{5pt}
\pgfsetendarrow{\pgfarrowto}
\pgfnodeconnline{node7}{node9}
\pgfnodelabel{node7}{node9}[0.5][0pt]{\pgfbox[center,center]{$Take_2$}}
\pgfsetendarrow{\pgfarrowto}
\pgfnodeconnline{node8}{node9}
\pgfnodelabel{node8}{node9}[0.5][0pt]{\pgfbox[center,center]{$Take_1$}}
\pgfsetendarrow{\pgfarrowto}
\pgfnodeconncurve{node9}{node1}{180}{180}{8cm}{10cm}
\pgfsetendarrow{\pgfarrowto}
\pgfnodeconncurve{node9}{node1}{0}{0}{8cm}{10cm}
\end{pgfpicture}
\caption{LTS Representation of Distilling Example 1}
\label{example1lts}
\end{figure}
Verification of Property 1 (mutual exclusion) fails for this transformed program; if the input event stream starts with 
$Request_1$, $Request_2$, $Take_1$, $Take_2$, $\ldots$, then the function calling sequence is $f_1$, $f_2$, $f_5$,
$f_7$, $f_9$, $\ldots$ and we can see that we end up in the function $f_9$, where both processes are using the critical resource.}
\end{example}
\begin{example}
\normalfont{The result of distilling Example 2 is shown in Figure \ref{example2distilled}, and the LTS representation of this program is shown in Figure \ref{example2lts}.
\begin{figure}[htb]
\hspace*{2cm}
\begin{tabular}{l}
$\expr{\where{\app{\var{f_1}}{\var{es}}}{
\fundef{\var{f_1}}{\abs{\var{es}}{\app{\app{\var{Cons}}{\var{(ObsState~T~T)}}}{\brackets{\smallcas{\var{es}}{\app{\app{\var{Cons}}{\var{e}}}{\var{es}}}{\longcas{\var{e}}{\var{Request_1}}{\app{\var{f_2}}{\var{es}}}{\var{Request_2}}{\app{\var{f_3}}{\var{es}}}{\var{\wildcard}}{\app{\var{f_1}}{\var{es}}}}}}}} \\
\fundef{\var{f_2}}{\abs{\var{es}}{\app{\app{\var{Cons}}{\var{(ObsState~W~T)}}}{\brackets{\smallcas{\var{es}}{\app{\app{\var{Cons}}{\var{e}}}{\var{es}}}{\longcas{\var{e}}{\var{Take_1}}{\app{\var{f_4}}{\var{es}}}{\var{Request_2}}{\app{\var{f_5}}{\var{es}}}{\var{\wildcard}}{\app{\var{f_2}}{\var{es}}}}}}}} \\
\fundef{\var{f_3}}{\abs{\var{es}}{\app{\app{\var{Cons}}{\var{(ObsState~T~W)}}}{\brackets{\smallcas{\var{es}}{\app{\app{\var{Cons}}{\var{e}}}{\var{es}}}{\longcas{\var{e}}{\var{Request_1}}{\app{\var{f_5}}{\var{es}}}{\var{Take_2}}{\app{\var{f_6}}{\var{es}}}{\var{\wildcard}}{\app{\var{f_3}}{\var{es}}}}}}}} \\
\fundef{\var{f_4}}{\abs{\var{es}}{\app{\app{\var{Cons}}{\var{(ObsState~U~T)}}}{\brackets{\smallcas{\var{es}}{\app{\app{\var{Cons}}{\var{e}}}{\var{es}}}{\cas{\var{e}}{\var{Release_1}}{\app{\var{f_1}}{\var{es}}}{\var{\wildcard}}{\app{\var{f_4}}{\var{es}}}}}}}} \\
\fundef{\var{f_5}}{\abs{\var{es}}{\app{\app{\var{Cons}}{\var{(ObsState~W~W)}}}{\brackets{\smallcas{\var{es}}{\app{\app{\var{Cons}}{\var{e}}}{\var{es}}}{\smallcas{\var{e}}{\var{\wildcard}}{\app{\var{f_5}}{\var{es}}}}}}}} \\
\fundef{\var{f_6}}{\abs{\var{es}}{\app{\app{\var{Cons}}{\var{(ObsState~T~U)}}}{\brackets{\smallcas{\var{es}}{\app{\app{\var{Cons}}{\var{e}}}{\var{es}}}{\cas{\var{e}}{\var{Release_2}}{\app{\var{f_1}}{\var{es}}}{\var{\wildcard}}{\app{\var{f_6}}{\var{es}}}}}}}}}}$
\end{tabular} 
\caption{Result of Distilling Example 2}
\label{example2distilled}
\end{figure}
\begin{figure}[htb]
\begin{pgfpicture}{0cm}{0cm}{20cm}{5.5cm}
\pgfnodebox{node1}[stroke]{\pgfxy(8.0,4.5)}{\parbox{1.1cm}{$f_1 \\ s_1 = T \\ s_2 = T$}}{5pt}{5pt}
\pgfnodebox{node2}[stroke]{\pgfxy(5.0,2.5)}{\parbox{1.1cm}{$f_2 \\ s_1 = W \\ s_2 = T$}}{5pt}{5pt}
\pgfsetendarrow{\pgfarrowto}
\pgfnodeconnline{node1}{node2}
\pgfnodelabel{node1}{node2}[0.5][0pt]{\pgfbox[center,center]{$Request_1$}}
\pgfnodebox{node3}[stroke]{\pgfxy(11.0,2.5)}{\parbox{1.1cm}{$f_3 \\ s_1 = T \\ s_2 = W$}}{5pt}{5pt}
\pgfsetendarrow{\pgfarrowto}
\pgfnodeconnline{node1}{node3}
\pgfnodelabel{node1}{node3}[0.5][0pt]{\pgfbox[center,center]{$Request_2$}}
\pgfnodebox{node4}[stroke]{\pgfxy(2.0,0.5)}{\parbox{1.1cm}{$f_4 \\ s_1 = U \\ s_2 = T$}}{5pt}{5pt}
\pgfsetendarrow{\pgfarrowto}
\pgfnodeconnline{node2}{node4}
\pgfnodelabel{node2}{node4}[0.5][0pt]{\pgfbox[center,center]{$Take_1$}}
\pgfsetendarrow{\pgfarrowto}
\pgfnodeconncurve{node4}{node1}{90}{180}{2cm}{2cm}
\pgfnodelabel{node4}{node1}[0.9][1cm]{\pgfbox[center,center]{$Release_1$}}
\pgfnodebox{node5}[stroke]{\pgfxy(8.0,0.5)}{\parbox{1.1cm}{$f_5 \\ s_1 = W \\ s_2 = W$}}{5pt}{5pt}
\pgfsetendarrow{\pgfarrowto}
\pgfnodeconnline{node2}{node5}
\pgfnodelabel{node2}{node5}[0.5][0pt]{\pgfbox[center,center]{$Request_2$}}
\pgfsetendarrow{\pgfarrowto}
\pgfnodeconnline{node3}{node5}
\pgfnodelabel{node3}{node5}[0.5][0pt]{\pgfbox[center,center]{$Request_1$}}
\pgfnodebox{node6}[stroke]{\pgfxy(14.0,0.5)}{\parbox{1.1cm}{$f_6 \\ s_1 = T \\ s_2 = U$}}{5pt}{5pt}
\pgfsetendarrow{\pgfarrowto}
\pgfnodeconnline{node3}{node6}
\pgfnodelabel{node3}{node6}[0.5][0pt]{\pgfbox[center,center]{$Take_2$}}
\pgfsetendarrow{\pgfarrowto}
\pgfnodeconncurve{node6}{node1}{90}{0}{2cm}{2cm}
\pgfnodelabel{node6}{node1}[0.9][-1cm]{\pgfbox[center,center]{$Release_2$}}
\end{pgfpicture}
\caption{LTS Representation of Distilling Example 2}
\label{example2lts}
\end{figure}
Verification of Property 1 (mutual exclusion) succeeds for this transformed program; we can easily see that there is no state in 
which both processes are using the critical resource. When trying to prove this property, as soon as we re-encounter 
any of the functions within the program, the value $True$ is returned by verification rule (6a).
However, verification of Property 2 (non-starvation) fails; if the input event stream starts with $Request_1$, $Request_2$, 
$\ldots$, then the function calling sequence is $f_1$, $f_2$, $f_5$, and we can see that we end up within the function $f_5$. 
At this point, both processes are waiting for the critical resource, so we need to prove that they will eventually get to use it. 
When trying to prove this eventuality property, we immediately re-encounter the function $f_5$, so the value $False$ is returned 
by verification rule (6b).}
\end{example}
\begin{example}
\normalfont{The result of distilling Example 3 is shown in Figure \ref{example3distilled}, and the LTS representation of this program is shown in Figure \ref{example3lts}.
\begin{figure}[htbp]
\hspace*{2cm}
\begin{tabular}{l}
$\expr{\where{\app{\var{f_1}}{\var{es}}}{
\fundef{\var{f_1}}{\abs{\var{es}}{\app{\app{\var{Cons}}{\var{(ObsState~T~T)}}}{\brackets{\smallcas{\var{es}}{\app{\app{\var{Cons}}{\var{e}}}{\var{es}}}{\longcas{\var{e}}{\var{Request_1}}{\app{\var{f_2}}{\var{es}}}{\var{Request_2}}{\app{\var{f_3}}{\var{es}}}{\var{\wildcard}}{\app{\var{f_1}}{\var{es}}}}}}}} \\
\fundef{\var{f_2}}{\abs{\var{es}}{\app{\app{\var{Cons}}{\var{(ObsState~W~T)}}}{\brackets{\smallcas{\var{es}}{\app{\app{\var{Cons}}{\var{e}}}{\var{es}}}{\longcas{\var{e}}{\var{Take_1}}{\app{\var{f_4}}{\var{es}}}{\var{Request_2}}{\app{\var{f_6}}{\var{es}}}{\var{\wildcard}}{\app{\var{f_2}}{\var{es}}}}}}}} \\
\fundef{\var{f_3}}{\abs{\var{es}}{\app{\app{\var{Cons}}{\var{(ObsState~T~W)}}}{\brackets{\smallcas{\var{es}}{\app{\app{\var{Cons}}{\var{e}}}{\var{es}}}{\longcas{\var{e}}{\var{Take_2}}{\app{\var{f_5}}{\var{es}}}{\var{Request_1}}{\app{\var{f_7}}{\var{es}}}{\var{\wildcard}}{\app{\var{f_3}}{\var{es}}}}}}}} \\
\fundef{\var{f_4}}{\abs{\var{es}}{\app{\app{\var{Cons}}{\var{(ObsState~U~T)}}}{\brackets{\smallcas{\var{es}}{\app{\app{\var{Cons}}{\var{e}}}{\var{es}}}{\longcas{\var{e}}{\var{Release_1}}{\app{\var{f_1}}{\var{es}}}{\var{Request_2}}{\app{\var{f_8}}{\var{es}}}{\var{\wildcard}}{\app{\var{f_4}}{\var{es}}}}}}}} \\
\fundef{\var{f_5}}{\abs{\var{es}}{\app{\app{\var{Cons}}{\var{(ObsState~T~U)}}}{\brackets{\smallcas{\var{es}}{\app{\app{\var{Cons}}{\var{e}}}{\var{es}}}{\longcas{\var{e}}{\var{Release_2}}{\app{\var{f_1}}{\var{es}}}{\var{Request_1}}{\app{\var{f_9}}{\var{es}}}{\var{\wildcard}}{\app{\var{f_5}}{\var{es}}}}}}}} \\
\fundef{\var{f_6}}{\abs{\var{es}}{\app{\app{\var{Cons}}{\var{(ObsState~W~W)}}}{\brackets{\smallcas{\var{es}}{\app{\app{\var{Cons}}{\var{e}}}{\var{es}}}{\cas{\var{e}}{\var{Take_1}}{\app{\var{f_8}}{\var{es}}}{\var{\wildcard}}{\app{\var{f_6}}{\var{es}}}}}}}} \\
\fundef{\var{f_7}}{\abs{\var{es}}{\app{\app{\var{Cons}}{\var{(ObsState~W~W)}}}{\brackets{\smallcas{\var{es}}{\app{\app{\var{Cons}}{\var{e}}}{\var{es}}}{\cas{\var{e}}{\var{Take_2}}{\app{\var{f_9}}{\var{es}}}{\var{\wildcard}}{\app{\var{f_7}}{\var{es}}}}}}}} \\
\fundef{\var{f_8}}{\abs{\var{es}}{\app{\app{\var{Cons}}{\var{(ObsState~U~W)}}}{\brackets{\smallcas{\var{es}}{\app{\app{\var{Cons}}{\var{e}}}{\var{es}}}{\cas{\var{e}}{\var{Release_1}}{\app{\var{f_3}}{\var{es}}}{\var{\wildcard}}{\app{\var{f_8}}{\var{es}}}}}}}} \\
\fundef{\var{f_9}}{\abs{\var{es}}{\app{\app{\var{Cons}}{\var{(ObsState~W~U)}}}{\brackets{\smallcas{\var{es}}{\app{\app{\var{Cons}}{\var{e}}}{\var{es}}}{\cas{\var{e}}{\var{Release_2}}{\app{\var{f_2}}{\var{es}}}{\var{\wildcard}}{\app{\var{f_9}}{\var{es}}}}}}}}}}$
\end{tabular} 
\caption{Result of Distilling Example 3}
\label{example3distilled}
\end{figure}
\begin{figure}[htb]
\begin{pgfpicture}{0cm}{1cm}{20cm}{10cm}
\pgfnodebox{node1}[stroke]{\pgfxy(8.0,9.5)}{\parbox{1.1cm}{$f_1 \\ s_1 = T \\ s_2 = T$}}{5pt}{5pt}
\pgfnodebox{node2}[stroke]{\pgfxy(5.0,7.5)}{\parbox{1.1cm}{$f_2 \\ s_1 = W \\ s_2 = T$}}{5pt}{5pt}
\pgfsetendarrow{\pgfarrowto}
\pgfnodeconnline{node1}{node2}
\pgfnodelabel{node1}{node2}[0.5][0pt]{\pgfbox[center,center]{$Request_1$}}
\pgfnodebox{node3}[stroke]{\pgfxy(11.0,7.5)}{\parbox{1.1cm}{$f_3 \\ s_1 = T \\ s_2 = W$}}{5pt}{5pt}
\pgfsetendarrow{\pgfarrowto}
\pgfnodeconnline{node1}{node3}
\pgfnodelabel{node1}{node3}[0.5][0pt]{\pgfbox[center,center]{$Request_2$}}
\pgfnodebox{node4}[stroke]{\pgfxy(2.0,4.5)}{\parbox{1.1cm}{$f_4 \\ s_1 = U \\ s_2 = T$}}{5pt}{5pt}
\pgfsetendarrow{\pgfarrowto}
\pgfnodeconnline{node2}{node4}
\pgfnodelabel{node2}{node4}[0.5][0pt]{\pgfbox[center,center]{$Take_1$}}
\pgfsetendarrow{\pgfarrowto}
\pgfnodeconncurve{node4}{node1}{90}{180}{2cm}{2cm}
\pgfnodelabel{node4}{node1}[0.9][1cm]{\pgfbox[center,center]{$Release_1$}}
\pgfnodebox{node6}[stroke]{\pgfxy(5.0,4.5)}{\parbox{1.1cm}{$f_6 \\ s_1 = W \\ s_2 = W$}}{5pt}{5pt}
\pgfsetendarrow{\pgfarrowto}
\pgfnodeconnline{node2}{node6}
\pgfnodelabel{node2}{node6}[0.5][0pt]{\pgfbox[center,center]{$Request_2$}}
\pgfnodebox{node7}[stroke]{\pgfxy(11.0,4.5)}{\parbox{1.1cm}{$f_7 \\ s_1 = W \\ s_2 = W$}}{5pt}{5pt}
\pgfsetendarrow{\pgfarrowto}
\pgfnodeconnline{node3}{node7}
\pgfnodelabel{node3}{node7}[0.5][0pt]{\pgfbox[center,center]{$Request_1$}}
\pgfnodebox{node5}[stroke]{\pgfxy(14.0,4.5)}{\parbox{1.1cm}{$f_5 \\ s_1 = T \\ s_2 = U$}}{5pt}{5pt}
\pgfsetendarrow{\pgfarrowto}
\pgfnodeconnline{node3}{node5}
\pgfnodelabel{node3}{node5}[0.5][0pt]{\pgfbox[center,center]{$Take_2$}}
\pgfsetendarrow{\pgfarrowto}
\pgfnodeconncurve{node5}{node1}{90}{0}{2cm}{2cm}
\pgfnodelabel{node5}{node1}[0.9][-1cm]{\pgfbox[center,center]{$Release_2$}}
\pgfnodebox{node8}[stroke]{\pgfxy(5.0,1.5)}{\parbox{1.1cm}{$f_8 \\ s_1 = U \\ s_2 = W$}}{5pt}{5pt}
\pgfsetendarrow{\pgfarrowto}
\pgfnodeconnline{node6}{node8}
\pgfnodelabel{node6}{node8}[0.5][0pt]{\pgfbox[center,center]{$Take_1$}}
\pgfsetendarrow{\pgfarrowto}
\pgfnodeconnline{node4}{node8}
\pgfnodelabel{node4}{node8}[0.5][0pt]{\pgfbox[center,center]{$Request_2$}}
\pgfsetendarrow{\pgfarrowto}
\pgfnodeconnline{node8}{node3}
\pgfnodelabel{node8}{node3}[0.3][0pt]{\pgfbox[center,center]{$Release_1$}}
\pgfnodebox{node9}[stroke]{\pgfxy(11.0,1.5)}{\parbox{1.1cm}{$f_9 \\ s_1 = W \\ s_2 = U$}}{5pt}{5pt}
\pgfsetendarrow{\pgfarrowto}
\pgfnodeconnline{node7}{node9}
\pgfnodelabel{node7}{node9}[0.5][0pt]{\pgfbox[center,center]{$Take_2$}}
\pgfsetendarrow{\pgfarrowto}
\pgfnodeconnline{node5}{node9}
\pgfnodelabel{node5}{node9}[0.5][0pt]{\pgfbox[center,center]{$Request_1$}}
\pgfsetendarrow{\pgfarrowto}
\pgfnodeconnline{node9}{node2}
\pgfnodelabel{node9}{node2}[0.3][0pt]{\pgfbox[center,center]{$Release_2$}}
\end{pgfpicture}
\caption{LTS Representation of Distilling Example 3}
\label{example3lts}
\end{figure}
We can see that the use of tickets is completely transformed away and that the resulting program has a finite number of states.
This is where distillation provides an advantage over other transformation techniques such as positive supercompilation
which are not able to remove as many intermediate data structures and thus to transform away the use of tickets.
Verification of both Property 1 (mutual exclusion) and Property 2 (non-starvation) succeed for this transformed program.
The proof of Property 1 is quite straightforward and similar to the proof of this property for Example 2. If we consider the 
proof of Property 2 for process 1, if the event $Request_1$ has just occurred, then we must be in one of the functions 
$f_2$, $f_7$ or $f_9$. There is a single exit from $f_7$ to $f_9$ by event $Take_2$, and a single exit from $f_9$ to path
$f_2$ by event $Release_2$. Thus, we must eventually end up in function $f_2$ after a $Request_1$ event. From $f_2$,
we must eventually end up in a state in which process 1 is using the critical resource, either directly by event $Take_1$, or
indirectly with event $Request_2$ preceding $Take_1$. The proof of Property 2 for process 2 is similar.
}
\end{example} 
\section{Conclusion and Related Work}

In this paper, we have shown how a fold/unfold program transformation technique can be used to verify 
both safety and liveness properties of reactive systems which have been specified using a functional language. 
Many corresponding techniques have been developed for verifying temporal properties for logic programs 
\cite{LEUSCHEL99,ROYCHOUDHURI00,FIORAVANTI01,PETTOROSSI09,SEKI11}). Some of these techniques
have been developed only for safety properties, while others can be used to verify both safety and liveness properties.
Due to the use of a different programming paradigm, it is difficult to compare the relative power of these techniques to our own.
However, we argue that the use of a more powerful program transformation algorithm will remove more intermediate data 
structures, and thus be capable of proving more properties directly within the same framework, without the need for making 
use of external solvers.

Very few techniques have been developed for verifying temporal properties for functional programs other than the work of 
Lisitsa and Nemytykh \cite{LISITSA07,LISITSA08}. Their approach uses supercompilation \cite{TURCHIN86,SORENSEN96} as 
the fold/unfold transformation methodology, where our own approach uses distillation \cite{HAMILTON07A,HAMILTON12}.
Since distillation has been shown to be more powerful than positive supercompilation, it follows that we should be able to
verify more properties using our approach (such as the properties we verify for Lamport's bakery algorithm in Example 3). 
Also, the work of Lisitsa and Nemytykh can verify only safety properties, while
our approach can be used to verify both safety and liveness properties.

One other area of work related to our own is the work on using Higher Order Recursion Schemes (HORS) to verify temporal
properties of functional programs. HORS are a kind of higher order tree grammar for generating a (potentially infinite) tree
and are well-suited to the purpose of verification since they have a decidable mu-calculus model checking problem, as proved
by Ong \cite{ONG06}. Kobayashi \cite{KOBAYASHI09} first showed how this approach can be used to verify safety properties 
of higher order functional programs. This approach was then extended to also verify liveness properties by Lester et al. 
\cite{LESTER10}. These approaches have a very bad worst-case time complexity, but techniques have been developed to 
ameliorate this to a certain extent. It does however appear likely that this approach will be able to verify more properties than 
our own approach but much less efficiently.

\section*{Acknowledgements}
This work was supported, in part, by Science Foundation Ireland grant 10/CE/I1855 to Lero - the Irish Software Engineering Research Centre (www.lero.ie), and by the School of Computing, Dublin City University.

\bibliographystyle{eptcs}

\bibliography{mybib}

\end{document}